\documentclass{emulateapj}

\newcommand{\be}{\begin{equation}}
\newcommand{\ee}{\end{equation}}
\newcommand{\ba}{\begin{eqnarray}}
\newcommand{\ea}{\end{eqnarray}}
\newcommand{\bfi}{\begin{figure}
\epsfxsize=9cm
\epsffile}
\newcommand{\efi}{\end{figure}}

\def\mnras{MNRAS}
\def\apjl{ApJL}
\def\apjs{ApJS}

\def\hmpci{\,h{\rm {Mpc}^{-1}}}
\def\mpc{\,h^{-1}{\rm {Mpc}}}
\def\kpc{\,h^{-1}{\rm {kpc}}}

\begin{document}

\title{The influence of baryons on the clustering of matter and weak lensing
  surveys} \author{Y. P. Jing$^{1,4}$, Pengjie Zhang$^{1,4}$, W.P.
  Lin$^{1,4}$, L. Gao$^{2,3}$, V. Springel$^{2}$ } \affil{$^1$ Shanghai
  Astronomical Observatory, Nandan Road 80, Shanghai, China} \affil{$^2$
  Max-Planck-Institut f\"ur Astrophysik, Karl-Schwarzschild-Strasse 1, 85748
  Garching, Germany} \affil{$^3$ Institute for Computational Cosmology,
  Physics Department, Durham. U.K.}  \affil{$^4$ Joint Institute for Galaxy
  and Cosmology (JOINGC) of SHAO and USTC} \affil{e-mail: ypjing@shao.ac.cn}

\begin{abstract}

  Future weak lensing measurements of cosmic shear will reach such
  high accuracy that second order effects in weak lensing modeling,
  like the influence of baryons on structure formation, become
  important.  We use a controlled set of high-resolution cosmological
  simulations to quantify this effect by comparing pure N-body dark
  matter runs with corresponding hydrodynamical simulations, carried
  out both in non-radiative, and in dissipative form with cooling and
  star formation.  In both hydrodynamical simulations, the clustering
  of the gas is suppressed while that of dark matter is boosted at
  scales $k>1\hmpci$. Despite this counterbalance between dark matter
  and gas, the clustering of the total matter is suppressed by up to 1
  percent at $1\la k \la 10\hmpci$, while for $k \approx 20\hmpci$ it is
  boosted, up to 2 percent in the non-radiative run and 10 percent in
  the run with star formation. The stellar mass formed in the latter
  is highly biased relative to the dark matter in the pure N-body
  simulation.  Using our power spectrum measurements to predict the
  effect of baryons on the weak lensing signal at $100<l<10000$, we
  find that baryons may change the lensing power spectrum by less than
  0.5 percent at $l<1000$, but by 1 to 10 percent at
  $1000<l<10000$. The size of the effect exceeds the predicted
  accuracy of future lensing power spectrum measurements and will
  likely be detected.  Precise determinations of cosmological
  parameters with weak lensing, and studies of small-scale
  fluctuations and clustering, therefore rely on properly including
  baryonic physics.
\end{abstract}


\section{Introduction}
Weak gravitational lensing directly measures the projected mass
distribution and is emerging as one of the most powerful and robust
probes of the large scale structure of the universe, and the nature of
dark matter, dark energy and gravity
\citep{Huterer02,Hu02,Jain03,Takada04,Ishak05,Knox05}. Ongoing and
upcoming surveys such 
as CFHTLS\footnote{http://www.cfht.hawaii.edu/Science/CFHLS},
DES\footnote{http://www.darkenergysurvey.org/},
LSST\footnote{http://www.lsst.org/},
Pan-STARRS\footnote{http://pan-starrs.ifa.hawaii.edu/},
SKA\footnote{http://www.skatelescope.org/} and
SNAP\footnote{http://snap.lbl.gov/} will significantly reduce
statistical errors in lensing power spectrum measurements to the
$<1\%$ level at $l\sim 1000$.  As new analysis techniques enable a
significant reduction of systematic errors in cosmic shear
\citep[][and references therein]{Jarvis04,Jain05,Heymans05} and cosmic
magnification measurements
\citep{Menard02,Scranton05,Zhang05a,Zhang05b}, weak lensing
measurement is quickly entering the precision era. To match this
accuracy, many simplifications in theoretical modeling have to be
scrutinized in detail
\citep{Bernardeau98,Schneider98,Dodelson05a,Dodelson05b,White05}.

A widely adopted simplification in weak lensing modeling is the
assumption that baryons trace dark matter perfectly. With this
simplification, weak lensing involves only gravity and collision-less
dark matter dynamics, allowing it to be accurately predicted with the
aid of N-body simulations.  However, on small scales, baryons {\em do
not follow} the dark matter distribution.  Recently, \citet{white04}
and \citet{zhanknox04} estimated the effects of cooling gas and
intra-cluster gas on the lensing power spectrum, respectively.  These
two components of baryons can both have an effect of a few percent on
the lensing power spectrum at $l\sim 3000$, but with opposite
signs. Such effects exceed future measurement errors and are certainly
relevant in order to exploit the full power of precision measurements
of weak lensing.  However, analytical models are simplified with {\it
ad hoc} parameters and lack the ability to deal with back-reactions of
baryons on the dark matter. They can not robustly answer several key
questions such as (1) to what level the two effects cancel, (2) at
what scales each of them dominates, and (3) how large the effect is
for the non-virialized intergalactic medium where most of the baryons
reside.  To quantify the effect of baryons on lensing statistics
accurately, it is hence necessary to use hydrodynamical simulations
with all relevant gas physics included.  In this {\it letter}, we analyze
a controlled set of simulations to address this issue.

\section{Baryonic effect on the clustering of cosmic matter}
\subsection{Simulations}
We use the {\small GADGET2} code \citep{springel01,springel05} to
simulate structure formation in a concordance cosmological model. The
cosmological and initial density fluctuation parameters of the model
are ($\Omega_m,\Omega_\Lambda,\Omega_b,\sigma_8,n,h$)=($0.268, 0.732,
0.044,0.85,1,0.71$). Three simulations were run with the same initial
fluctuations for a cubic box of 100 $\mpc$. The first simulation is a
pure dark matter cosmological simulation. The second one is a gas
simulation where no radiative cooling of gas is considered. The last
run is a gas simulation that includes the physical processes of
radiative cooling and star formation.  It also includes supernova
feedback, outflows by galactic winds, and a sub-resolution multiphase
model for the interstellar medium as detailed in \citet{sh03}. In all
simulations, we use $512^3$ particles to represent each component of
dark matter and gas.

Uncertainties in simulating star formation could bias our
  study. To check the robustness of our star formation 
  run, we measured the stellar mass density $\Omega_\star$ in
  units of the critical density at $z=0$.   
We found 
$\Omega_\star=0.0034$.  This is in
reasonable agreement with recent observational results (e.g.
\citet[][$\Omega_\star=0.0035$]{fhp98}, \citet[][$\Omega_\star=0.0028\pm
0.0008$]{bell}). The star formation history of the simulation is therefore
close enough to reality for the purposes of the present paper.

\subsection{Clustering in the non-radiative run}
We measure the mass power spectrum $P(k)$ for matter, gas and stars
separately, as well as for the total matter density in the simulations.  The
Triangular Shaped Cloud (TSC) interpolation function $W(r)$ is used to obtain
the smoothed density field on a uniform grid of $1024^3$ cells, and the power
spectrum is determined from the Fourier transform of the density field.  We
use the method of \citet{jing2005} to correct for the smoothing and aliasing
effects caused by the TSC mass interpolation and for the noise caused by the
particle discreteness.

The power spectrum of each matter component is plotted in Figure 1 for
redshifts $z=0$ and $z=1$. Here we use the mass variance $\Delta^2(k)$
per logarithmic interval in wavelength which is related to $P(k)$ by
$\Delta^2(k)=4\pi k^3 P(k)/(2\pi)^3$. According to the extensive tests
of \citet{hou05}, the power spectrum is affected by the force
resolution at $k>k_{\eta} \approx 0.3\times 2\pi/\eta$ , where $\eta$
is the Plummer-equivalent force softening length used in the
simulation. In our simulations, $\eta=5\kpc$ for the pure dark matter
and non-radiative runs, and $\eta=9\kpc$ for the star formation
run. Down to the scale of $k\approx 20 \hmpci$, the force resolution
should have negligible effects.

Comparing the results of the pure dark matter and the non-radiative runs, one
finds that the gas has a weaker clustering than the dark matter on small
scales ($k>1\hmpci$). This can be interpreted as the result of gas pressure
which reduces small-scale structure in the gas distribution.
\citet{zhanknox04} approximated the hot baryon distribution by assuming that
the gas is in hydrostatic equilibrium in NFW dark halos \citep{makino98}, and
found a qualitatively similar result. But quantitatively, the gas clustering
suppression in their simple analytic estimate is stronger than our simulations
indicate. At $k=10\hmpci$ and $z=0$, the $\Delta^2(k)$ of gas is 25 percent
lower than that of dark matter in our non-radiative simulation (see Figure
\ref{psdiff}), while the $\Delta^2(k)$ of gas is 50 percent lower in their
analytic estimate. We also see a more extended baryonic effect at $k<1\hmpci$
which may be caused by filaments and is beyond the exploration of their halo
model. These effects indicate the need for hydrodynamical simulations to
accurately assess the effect of baryons on the clustering of cosmic matter.

An interesting result from our non-radiative simulation is that its
dark matter has a stronger clustering than found in the pure dark
matter run. To show the difference clearly, we plot it separately in
Figure \ref{psdiff}, where we also show the clustering differences of
other matter components relative to the pure dark matter run.  The
dark matter clustering is a few percent stronger in the non-radiative
cooling simulation at $k >1\hmpci$, and is more than 10\% stronger at
$k >10\hmpci$. This behavior can be understood as a result of the
gravitational back reaction due to hot gas, since the gas is hotter
than the dark matter virial temperature in the central regions of dark
halos, due to its collisional nature \citep[see,][]{tormen, linwp}.

\bfi{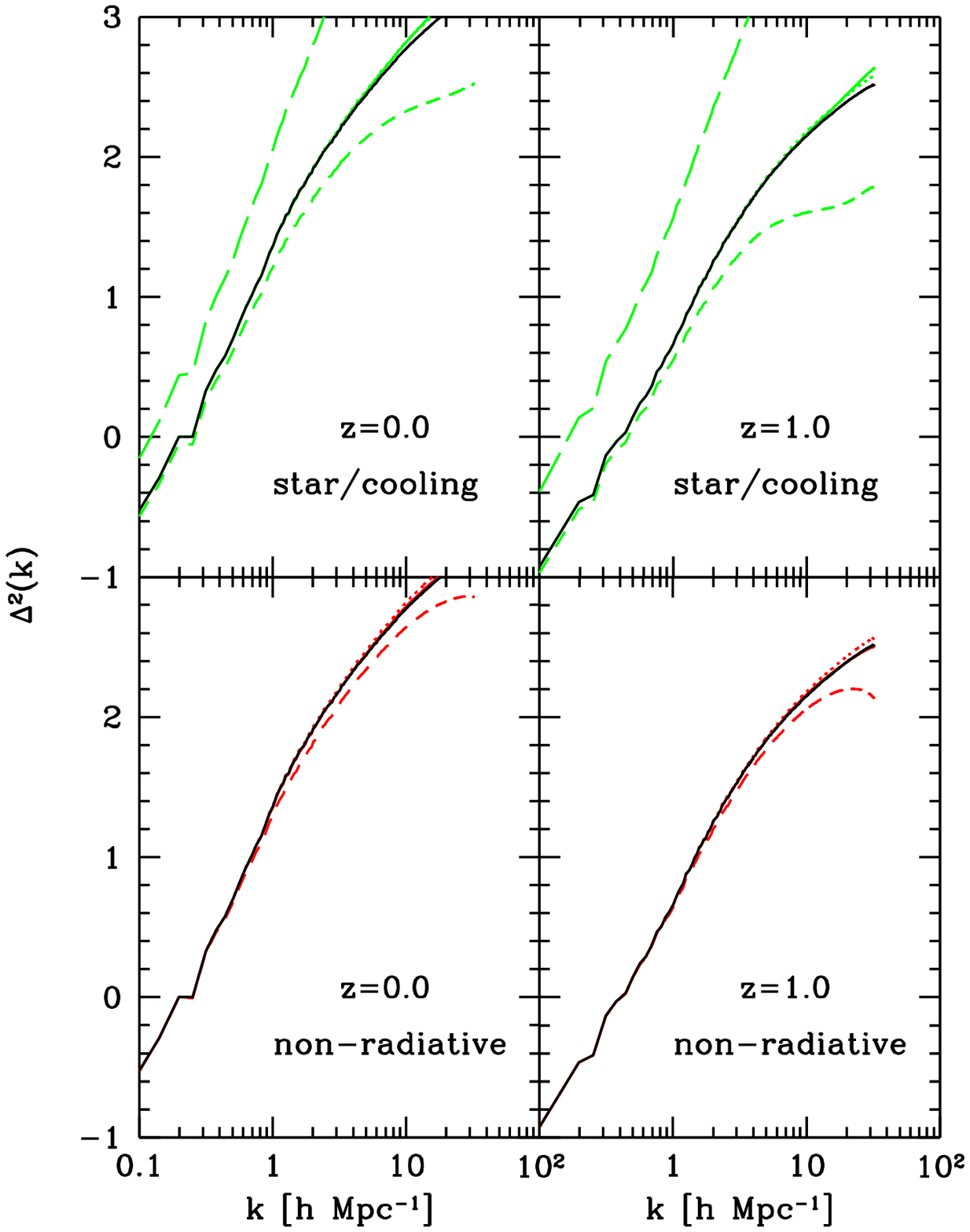} 
\caption{The power spectrum of density fluctuations of each matter
component in the non-radiative gas simulation (bottom panels), and in
the gas cooling and star formation simulation (top panels), at
redshifts $z=1$ and $z=0$. The results are compared with the spectrum
of the pure dark matter simulation (black solid lines). The colored
dotted lines, 
dashed lines, long dashed lines, and solid lines are 
plotted for the dark matter, gas, stellar and total matter density
field, respectively.}
\label{ps}
\efi

Compared to the power spectra of gas and dark matter, the $P_{\rm
NR}^{\rm tot}(k)$ of the total matter distribution in the
non-radiative run differs less dramatically from the $P_{\rm DM}(k)$
of the pure dark matter run\footnote{Here we use the subscripts DM,
NR, SF for the pure dark matter, non-radiative, and star formation
runs, respectively, and the superscripts dm, gas, star, and tot for
each matter component.}, because the baryonic suppression of the gas
clustering is partially counterbalanced by its effect on the dark
matter. However, there still exists a significant difference in the
total matter clustering between the two runs (see Figures 1 and
2). The total matter spectrum $P_{\rm NR}^{\rm tot}(k)$ is a few
percent lower than $P_{\rm DM}(k)$ at $k \approx 2\hmpci$, and is a
few percent higher than $P_{\rm DM}(k)$ at $k \approx 10\hmpci$. At
the smaller scales, the difference becomes smaller because $P_{\rm
NR}^{\rm gas}(k)$ drops faster than $P_{\rm NR}^{\rm dm}(k)$
increases. Although the formal gravitational resolution limit of the
simulation is $k\approx 40\hmpci$, hydrodynamical convergence can only
be expected on somewhat larger scales, thus it will be interesting to
check the results at $k>10\hmpci$ shown in Figure 2 with higher
resolution simulations.

\subsection{Clustering in the simulation with star formation }
In our star formation run, the stellar matter is found to be
significantly more clustered than dark matter or gas (Figure 1). Its
power spectrum $P_{\rm SF}^{\rm star}(k)$ is a few times higher than
the dark matter counterpart $P_{\rm SF}^{\rm dm}(k)$, and the bias
factor increases towards smaller scales. This result is expected,
because the stars can form only in high density regions and thus are
highly biased tracers of the matter distribution.

\bfi{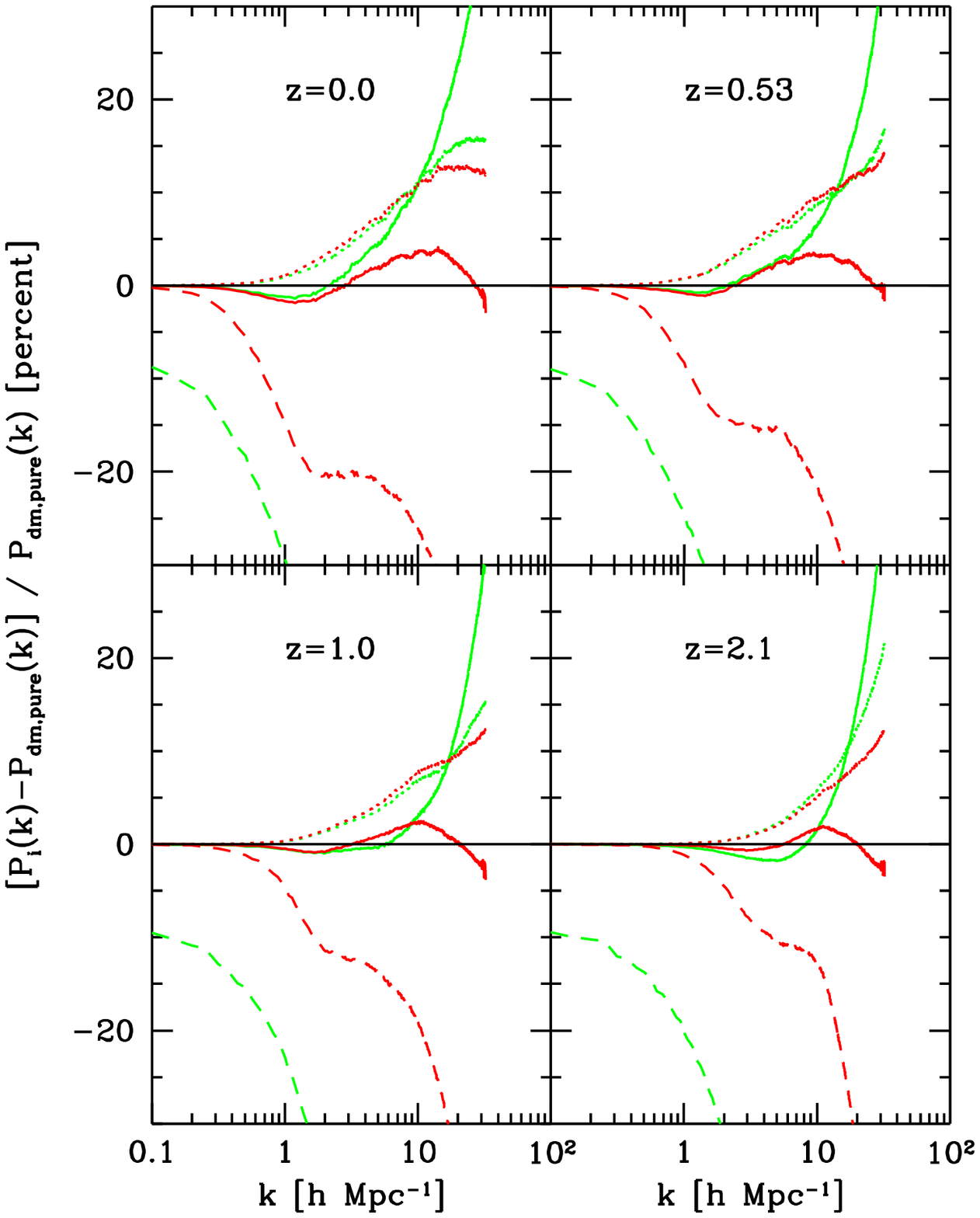} 
\caption{The influence of baryons on the clustering of each matter
component. The plot shows the relative difference (in percent) of the
power spectrum of each matter component in the gas simulations
relative to the pure dark matter simulation. The results of the
non-radiative run are plotted with red color, and those of the star
formation run with green colors. The colored dotted lines, dashed lines,
and solid lines are plotted for the dark matter and gaseous
components, and the total matter density field, respectively. The
stellar component in the star formation run is omitted here, as the
difference can be easily infered from Figure~\ref{ps}. }
\label{psdiff}
\efi

On the other hand, the power spectrum of gas $P_{\rm SF}^{\rm gas}(k)$
is significantly lower than $P_{\rm SF}^{\rm dm}(k)$. Apart from the
mild effects of shock heating and gas pressure that can be infered
from the non-radiative simulation, the main reason is that gas is
under-represented relative to the dark matter in high density regions,
where part of the gas has cooled and turned into stars. Because of
this anti-bias, $P_{\rm SF}^{\rm gas}(k)$ is always smaller by $\sim
10\%$ than $P_{\rm SF}^{\rm dm}(k)$ even on large scales (Figure 2),
and drops much faster at small scales than the counterpart $P_{\rm
NR}^{\rm gas}(k)$ in the non-radiative run.  The power spectrum of
dark matter $P_{\rm SF}^{\rm dm}(k)$ looks very similar to the
counterpart $P_{\rm NR}^{\rm dm}(k)$ in the non-radiative run, except
the former is higher at $k>10\hmpci$ because of the baryon
condensation in the star formation run.

The power spectrum of the total matter density $P_{\rm SF}^{\rm
tot}(k)$ shows a similar decrease at $k\la 2\hmpci$ as found in the
non-radiative run. As we discussed before, this behavior is mainly
due to shock heating and the thermal pressure of the gas. This is
confirmed in our star formation run, indicating that this feature is
robust against the star formation processes. The $P_{\rm SF}^{\rm
tot}(k)$ is higher than $P_{\rm NR}^{\rm tot}(k)$ at $k>10 \hmpci$ at
$z<1$, which is mainly caused by the baryon condensation due to gas
cooling. We also note that $P_{\rm SF}^{\rm tot}(k)$ is slightly lower
at $k \approx 8 \hmpci$ than the counterparts $P_{\rm NR}^{\rm
tot}(k)$ and $P_{\rm DM}(k)$ at $z\ga 1$, which likely results from
the feedback heating due to active star formation at high redshift.

\section{The baryonic effect on weak lensing}
In this section, we study the influence of baryons on the weak lensing
shear power spectrum $C_l$. We use Limber's approximation
\citep{Limber54} to calculate $C_l$ from the simulated $P^{\rm tot}$. For
a flat universe, $C_l$ and $P^{\rm tot}$ are related by
\begin{equation}
C_l=\left(\frac{3\Omega_m H_0^2}{2c^2}\right)^2\int P^{\rm
  tot}\left(\frac{l}{\chi},z\right)W^2(\chi,\chi_s)\chi^{-2} {\rm d}\chi\ ,
\end{equation}
where $W(\chi,\chi_s)=(1+z)\chi(1-\chi/\chi_s)$ is the lensing kernel,
and $\chi_s$ and $\chi$ are the comoving angular diameter distances to
the source and lens, respectively.

In Figure \ref{cldiff}, we show the effect of the baryons on the weak
lensing power spectrum by assuming that the lensed sources are at
redshifts $z_s=0.6$, 1.0, and 1.5, respectively. For the non-radiative
run, the lensing power signal is suppressed by less than one percent
at $100<l<1000$ and is then enhanced to about 1\% at $l=4000$. The
detailed behavior depends on the source redshift, and the relative
change of $C_l$ increases with the decrease of $z_s$. This is expected
since the baryons have more significant effects at lower redshifts
(Figure \ref{psdiff}).

Including more realistic star formation processes compensates for the
lensing power suppression at $100<l<1000$ seen in the non-radiative
run. As a result, the change of the lensing power by the presence of
baryons is smaller than $0.5\%$ for $z_s>0.6$ and $100<l<1000$. But at
$1000<l<10000$, the baryons can increase $C_l$ by up to 10 percent
depending on $z_s$.  Again, the relative change of $C_l$ increases
with the decrease of $z_s$.

\bfi{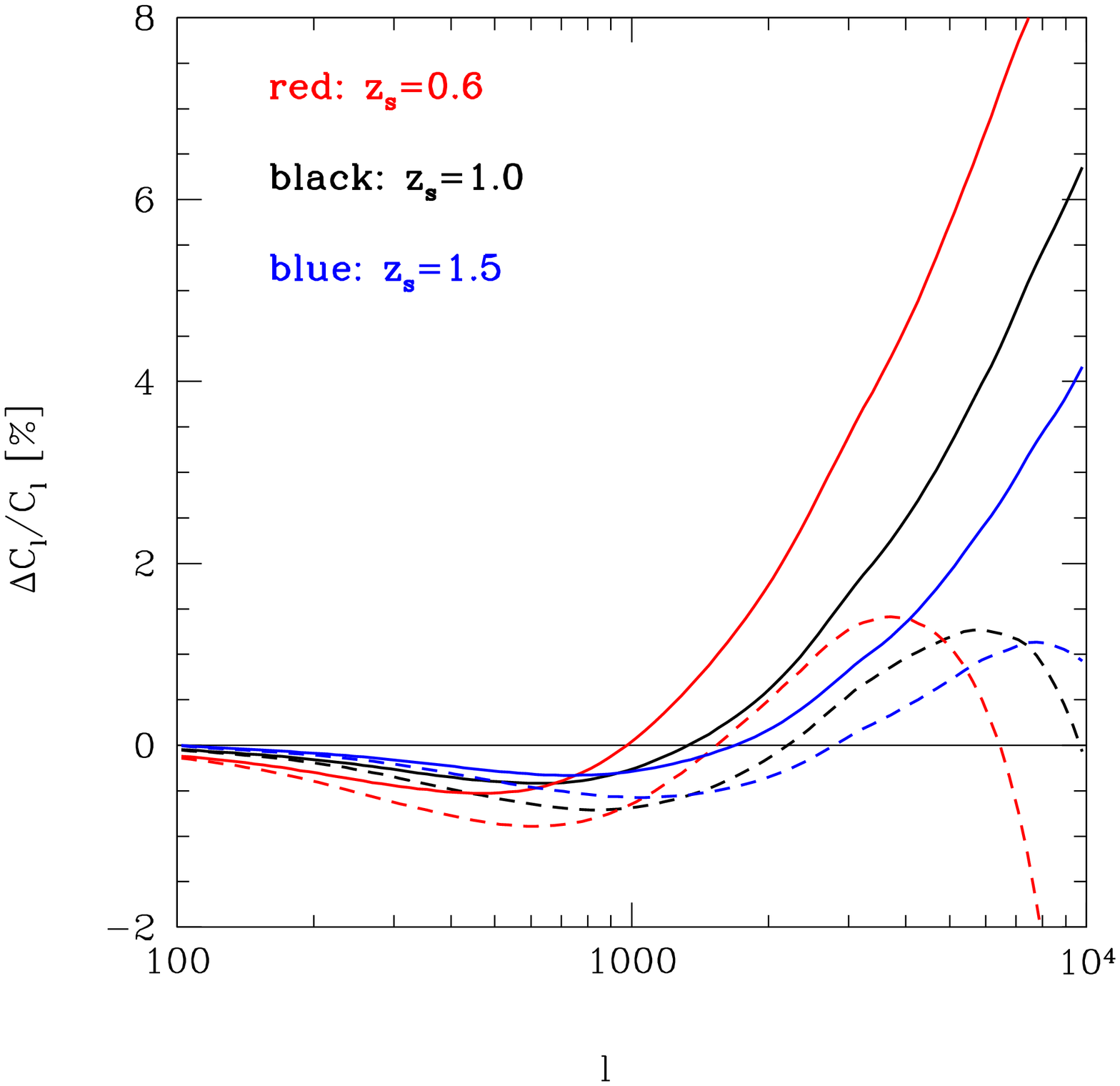} 
\caption{The effect of baryons on the shear power spectrum $C_l$,
expressed as relative difference (in percent) to the pure dark matter
model. Here the source galaxies are assumed to be at redshifts
$z_s=0.6$, $1.0$ and $1.5$, respectively. The dashed lines are the
predictions from our non-radiative run, while the solid lines give the
results for the star formation run. }
\label{cldiff}
\efi

Compared to the previous study of \citet{zhanknox04} for the effect of
hot baryons, our prediction for $\Delta C_l/C_l$ in the non-radiative
run is quite different from their finding of a monotonic decline of
$\Delta C_l/C_l$ with $l$ at $l>1000$. We argue that the main reason
for this is that they neglected the back reaction of the thermal
baryons on the dark matter in their analytical modeling.

To match the accuracy of future lensing surveys, matter power spectrum
at scales discussed in this {\it Letter} must be calibrated to $\sim
1\%$ accuracy \citep{Huterer05a}. Thus, the baryonic effect is
non-negligible for future lensing analysis. Here, we do a simple estimation to
demonstrate this point. 
The statistical errors of lensing power spectrum measurements
(assuming Gaussianity) are 
\be 
\frac{\Delta C_l}{C_l}=1\% \,
\left[\frac{l}{1000}\frac{\Delta l}{100}\frac{f_{\rm
sky}}{0.1}\right]^{-1/2}\left(1+\frac{\gamma^2_{\rm
rms}}{\bar{n}_gC_l}\right) , 
\ee where $n_g$ is the galaxy number
density, with typical value $\la 100/{\rm arcmin}^2$, and $\gamma_{\rm
rms}\sim 0.2$ is the rms fluctuation caused by the galaxy
ellipticities. At $l\la 3000$, the shot noise term is sub-dominant.
For future lensing missions with fractional sky coverage $f_{\rm
sky}\ga 0.1$, changes in $C_l$ caused by baryons are comparable to, or
larger than statistical errors, at $l\ga 1000$. To constrain cosmology
using measured $C_l$'s at all accessible scales ($l$ less than several
$10^4$), the effects due to baryons have to be taken into
account. Otherwise the derived constraints can be biased. For example,
since $C_l\propto \Omega^{\sim 1.2}_m \sigma_8^2$
\citep[e.g.][]{Van01}, neglecting this effect can lead to an
overestimate of $\Omega_m^{\sim 0.6}\sigma_8$ by several percent.
Also, it can cause an overestimate of the initial power index $n$. To
avoid such biases, an accurate modeling of the baryonic effects using
hydrodynamical simulations is required.

On the other hand, a precision measurement of the lensing signal at
$l$ of about a few thousand could be used to observationally determine
the baryonic effects and to constrain the galaxy formation
process. This can be done in two steps. One first uses the lensing
power spectrum at $l\la 1000$ to constrain $\Omega_m$, $\Omega_{\rm
DE}, \sigma_8$, etc. Even though this uses only part of the
information in the lensing power spectrum, the cosmological
constraints obtained are not significantly degraded
\citep{Huterer05b}. The derived cosmological model is then used to
predict $C_l$ (and its statistical fluctuations) at $l\ga 1000$,
assuming no baryonic effects.  One can then quantify the deviation of
the measured $C_l$ at $l\ga 1000$ from the predicted $C_l$.

\section{Conclusions}
In this {\it Letter}, we used a controlled set of high resolution
N-body and hydrodynamical/N-body simulations to study the influence of
baryons on the clustering of cosmic matter.
Since the three simulations we used have identical initial
  condition, simulated power spectra suffer essentially the same
  sample variance. Since we quantify the baryonic effect as 
  ratios of corresponding power spectra, the result presented in this
  {\it Letter} is 
  effectively free of sample variance. 
 In both the non-radiative
simulation and the simulation with gas cooling and star formation, the
clustering of the gas is suppressed while that of dark matter is
boosted at $k>1\hmpci$. The stellar mass is highly biased relative to
the dark matter in the pure N-body simulation.  Despite a partial
counterbalance between the dark matter and the gas, the clustering of
the total matter is suppressed by up to 1\% at $1\la k \la 10\hmpci$, and is
boosted up to 2\% in the non-radiative run, and 10\% in the star
formation run at $k \approx 20\hmpci$. Using these power spectrum
measurements to study the baryonic effect on the weak lensing shear
measurement at $100<l<10000$, we found that baryons can change the
shear power spectrum by less than 0.5\% at $l<1000$, but by 1\% to
10\% at $1000<l<10000$. Therefore, the influence of baryons on the
clustering of cosmic matter will be detected in future weak lensing
surveys. Understanding these baryonic effects is not only important
for galaxy formation, but also crucial for accurately determining
cosmological parameters with cosmic shear, and for constraining the
initial fluctuations on small scales.

\acknowledgments The simulations were run at Shanghai Supercomputer
Center.  The work at Shanghai is supported by the grants from NSFC
(Nos. 10125314, 10373012, 10533030) and from Shanghai Key Projects in
Basic research (No. 04JC14079 and 05XD14019). PJZ is supported by the
{\it One Hundred Talents} project of Chinese academy of science.


\begin{thebibliography}{}
\bibitem[Bell et al.(2003)]{bell} Bell, E.~F., McIntosh, D.~H., Katz, N., \& Weinberg, M.~D.\ 2003, \apjs, 149, 289 

\bibitem[Bernardeau(1998)]{Bernardeau98} Bernardeau, F.\ 1998, \aap, 
338, 375 

\bibitem[Dodelson \& Zhang(2005)]{Dodelson05a} Dodelson, S., \& 
Zhang, P.\ 2005, \prd, 72, 083001 

\bibitem[Dodelson et al.(2005)]{Dodelson05b} Dodelson, S., Shapiro,
  C. \& White, M.\ 2005, astro-ph/0508296

\bibitem[Fukugita et al.(1998)]{fhp98} Fukugita, M., Hogan, 
C.~J., \& Peebles, P.~J.~E.\ 1998, \apj, 503, 518 

\bibitem[Heymans et al.(2005)]{Heymans05} Heymans, C., et al. 2005,
  astro-ph/0506112

\bibitem[Hou et al.(2005)]{hou05} Hou, Y.~H., Jing, Y.~P., 
Zhao, D.~H., {B\"o}rner, G.\ 2005, \apj, 619, 667 

\bibitem[Hu(2002)]{Hu02} Hu, W.\ 2002, \prd, 66, 083515 

\bibitem[Huterer(2002)]{Huterer02}  Huterer,D. \  2002, \prd, 65,
  063001

\bibitem[Huterer \& Takada(2005)]{Huterer05a} Huterer, D., \& Takada,
  M. \ 2005, Astropart.Phys., 23, 369

\bibitem[Huterer \& White(2005)]{Huterer05b} Huterer, D., \& 
White, M.\ 2005, \prd, 72, 043002 

\bibitem[Ishak et al.(2005)]{Ishak05} Ishak, M., Upadhye, A., 
\& Spergel, D.~N.\ 2005, ArXiv Astrophysics e-prints, 
arXiv:astro-ph/0507184 

\bibitem[Jain \& Taylor(2003)]{Jain03} Jain, B., \& Taylor, 
A.\ 2003, Physical Review Letters, 91, 141302 

\bibitem[Jain et al.(2005)]{Jain05} Jain, B., Jarvis, M. \& Bernstein,
  G., 2005, astro-ph/0510231

\bibitem[Jarvis \& Jain(2004)]{Jarvis04} Jarvis, M. \&  Jain, B.,
  2004, astro-ph/0412234 

\bibitem[Jing(2005)]{jing2005} Jing, Y.~P.\ 2005, \apj, 620, 559 

\bibitem[Knox et al.(2005)]{Knox05} Knox, L., Song, Y.~-., \& 
Tyson, J.~A.\ 2005, ArXiv Astrophysics e-prints,
arXiv:astro-ph/0503644
  
\bibitem[Limber(1954)]{Limber54} Limber, D.~N.\ 1954, \apj, 119, 
655

\bibitem[Lin et al.(2006)]{linwp} Lin, W.P. et al., in preparation.

\bibitem[Makino et al.(1998)]{makino98} Makino, N., Sasaki, S., 
\& Suto, Y.\ 1998, \apj, 497, 555 

\bibitem[M{\'e}nard \& Bartelmann(2002)]{Menard02} M{\'e}nard, 
B., \& Bartelmann, M.\ 2002, \aap, 386, 784

\bibitem[Rasia et al.(2004)]{tormen} Rasia, E., Tormen, G., \& 
Moscardini, L.\ 2004, \mnras, 351, 237 

\bibitem[Schneider et al.(1998)]{Schneider98} Schneider, P., van 
Waerbeke, L., Jain, B., \& Kruse, G.\ 1998, \mnras, 296, 873 
 

\bibitem[Scranton et al.(2005)]{Scranton05} Scranton, R., et al.\ 
2005, \apj, 633, 589 

\bibitem[Springel(2005)]{springel05} Springel, V.\ 2005, \mnras, 
999 
 
\bibitem[Springel \& Hernquist(2003)]{sh03} Springel, V., \& 
Hernquist, L.\ 2003, \mnras, 339, 289 
 
\bibitem[Springel et al.(2001)]{springel01} Springel, V., Yoshida, 
N., \& White, S.~D.~M.\ 2001, New Astronomy, 6, 79 

\bibitem[Takada \& Jain(2004)]{Takada04} Takada, M., \& Jain, 
B.\ 2004, \mnras, 348, 897 

\bibitem[Van Waerbeke et al.(2001)]{Van01} Van Waerbeke, L., 
et al.\ 2001, \aap, 374, 757

\bibitem[White(2004)]{white04} White, M.\ 2004, Astroparticle 
Physics, 22, 211 

\bibitem[White(2005)]{White05} White, M.\ 2005, Astroparticle 
Physics, 23, 349 

\bibitem[Zhan \& Knox(2004)]{zhanknox04} Zhan, H., \& Knox, L.\ 
2004, \apjl, 616, L75 

\bibitem[Zhang \& Pen(2005a)]{Zhang05a}  Zhang, P.J. \& Pen, U.L.,
 2006, MNRAS, in press. astro-ph/0504551
\bibitem[Zhang \& Pen(2005b)]{Zhang05b} Zhang, P.J. \& Pen, U.L., 2005,
  \prl, 95, 241302


\end{thebibliography}
\end{document}